\documentclass[english,pra,preprint,aps,floats]{revtex4}
\usepackage[T1]{fontenc}
\usepackage[latin1]{inputenc}
\usepackage{graphicx}
\usepackage{amsmath}
\usepackage{babel}

\begin{document}

\title{A Modified Version  of the Waxman Algorithm}

\author{W. A. Berger\thanks{E-mail wb@wberger.com
} and H. G. Miller\thanks{E-Mail: hmiller@maple.up.ac.za
}}

\affiliation{Department of Physics, University of Pretoria, Pretoria 0002, South
Africa}

\begin{abstract}
The iterative algorithm recently proposed by Waxman for solving eigenvalue
problems, which relies on the method of moments,
has been  modified to improve 
its  convergence considerably without  sacrificing its benefits or elegance. The suggested modification is based on methods
to calculate low-lying eigenpairs of large bounded hermitian operators or matrices.

\noindent PACS\ 03.65.Ge,\ 02.60.Lj 
\end{abstract}
\maketitle
Recently, Waxman \cite{W98} has proposed a convergent iterative algorithm
for obtaining solutions of the eigenvalue problem, which does not
involve a matrix diagonalization. For operators which possess a continuum
as well as a set of bound states this is most advantageous \cite{AMP06}.
In the case of the ground state, for example, the eigenenergy, $\epsilon$,
is determined numerically as a function of the coupling constant of
the potential, $\lambda$, and inverted to yield the $\epsilon$ corresponding
to the required value of $\lambda$. The convergence rate of the algorithm,
therefore, depends on two factors: the number of iterations required
to find an eigensolution for a particular choice of $\epsilon$ and
the number of times this must be repeated in order to determine theThe iterative algorithm recently proposed by Waxman for solving eigenvalue
problems, which relies on the method of moments,
has been  modified to improve 
its  convergence considerably without  sacrificing its 
value of $\epsilon$ which corresponds to the desired value of $\lambda$.
In a recent paper we have shown that for many non-singular symmetric
potentials which vanish asymptotically, a simple analytical relationship
between the coupling constant of the potential and the ground state
eigenvalue exists  which can be used to reduce the number of times
$\lambda$ has to be calculated for a given value of  $\epsilon$ \cite{BM06}. Here we  show, that it is possible also to reduce the number of iterations necessary 
to determine an eigensolution for a particular choice of $\epsilon$.  Furthermore the existence of a simple analytical relationship between $\epsilon$ and $\lambda$ can be used to handle problematic situations that are referred to as pseudo convergence in other methods \cite{MB79} and which can occur in the Waxman algorithm as well. 

In the Waxman algorithm, eigenpairs are determined as functions of
the strength of the potential in the following manner \cite{W98}.
 Here we shall use the  abstract Hilbert Space notation for the most part 
and point out their meaning in a one dimensional coordinate space where
appropriate. Starting from \begin{equation}
(T-\lambda V)|u>=\epsilon|u>\label{heq}
\end{equation}
with
\begin{equation}
u(x)=<x|u>,\:\lim_{|x|->\infty}u(x)=0
\end{equation}
 where T is the kinetic energy operator (or  more generally  a suitable
hermitian operator for the unperturbed system); $\lambda>0$
is the strength parameter of the potential ($\lambda$V $>$ 0 and
V(x)$\rightarrow0\,\, as\,\,|x|\rightarrow\infty$) and the energy
eigenvalue, $\epsilon$ (with $\epsilon<0$), is negative and corresponds
to a bound state. Using Green's method a solution to equation (\ref{heq})
is given by \begin{equation}
|u>=\lambda G_{\epsilon}V|u>\label{G}\end{equation}
 where the Green's operator, $G_{\epsilon}$, can formally be defined
as \begin{equation}
G_{\epsilon}=(T-\epsilon)^{-1}
\end{equation}
and the corresponding Green's function satisfies
\begin{equation}
\lim_{|x|->\infty}G_{\epsilon}(x)=0.
\end{equation}
 Normalizing $|u>$ with a suitable reference state $|ref>$\begin{equation}
<ref|u>=1\end{equation}
 allows $\lambda$ to be written as (see equation (\ref{G})) \begin{equation}
\lambda=<ref|G_{\epsilon}V|u>^{-1}\label{l}\end{equation}
 which can then be used to eliminate $\lambda$ from equation (\ref{G})
\begin{equation}
|u>=\frac{G_{\epsilon}V|u>}{<ref|G_{\epsilon}V|u>}.\label{u}\end{equation}
 From equations (\ref{l}) and (\ref{u}), $\lambda$ can be determined
as a function of $\epsilon$ in the following manner. For a particular
choice of $\epsilon$ equation (\ref{u}) can be iterated \begin{equation}
|n+1>=\frac{G_{\epsilon}V|n>}{<ref|G_{\epsilon}V|n>}\label{iu}\end{equation}
 until it converges and $\lambda$ can then be determined from equation (\ref{l}).
Repeating for different values of $\epsilon$ yields a set of different
values of the potential strength $\lambda$. When enough points have
been determined, a simple interpolation procedure can be used to determine
the dependence of $\epsilon$ on $\lambda$.

If we assume, that u is square integrable and therefore can be  normalized we can choose $|u>$ as a
reference vector in (\ref{u}) yielding 
\begin{equation}
|u><u|G_{\epsilon}V|u>=G_{\epsilon}V|u>\label{laml}
\end{equation}
 from which  it is clearly seen, that we are dealing with an eigenvalue
problem for the operator $G_{\epsilon}V$ with eigenvalue $\lambda^{-1}=<u|G_{\epsilon}V|u>$.
The  equation corresponding to (\ref{iu}) is now given by \begin{equation}
|n+1>=\frac{G_{\epsilon}V|n>}{<n|G_{\epsilon}V|n>}.\label{ua}
\end{equation}
From (\ref{ua}) it immediately follows $<n|n+1>=1$ and therefore
  \begin{equation}
   |n+1>=|n>+c_{n}^{\bot}|n_{\bot}>
   \end{equation}
   where $<n|n_{\bot}>=0$, $<n_{\bot}|n_{\bot}>=1$ and $c_{n}^{\bot}\neq0$ as long as $|n>$ is not an eigenvector. Normalizing $|n+1>$  yields

\begin{equation}
|n+1>^{'}=(1+(c_{n}^{\perp})^{2})^{\frac{-1}{2}}(|n>+c_{n}^{\perp}|n_{\perp}>)\label{un}.
\end{equation}
 
Equation (\ref{un}) defines the start vector for the next iteration step and it can be easily seen, that the relative contribution of $|n>$ is reduced in each iteration step.
 It is now essential to note, that in the n+1 iteration step we have to deal
with two vectors: $|n>$ and $|n_{\perp}>$ which define (in
general) a two dimensional subspace. Therefore one could ask, does equation (\ref{un}) already define the best choice within this subspace to achieve optimal convergence. Remember  we are looking for a linear combination of the form 
\begin{equation}
|n+1>^{''}=c_{1}|n>+c_{2}G_{\epsilon}V|n>=d_{n}|n>+d_{n}^{\perp}|n_{\perp}>
\end{equation}
where the c's and the d's are chosen to satisfy the normalization
and the convergence requirements. Clearly, in the original Waxman scheme
 $c_{1}=0.$ This corresponds to using  the method of moments
or power method \cite{FF64,V62} to solve equation (\ref{laml}).

Another choice would be to diagonalize $G_{\epsilon}V$ projected onto the 2-dimensional subspace spanned by $|n>$ and $|n_{\perp}>$ and take one of the two eigenvectors  as the  new start vector in the next iteration step. Such an approach
has been proposed to calculate low-lying eigenvalues of an hermitian
operator \cite{BMK77} and generalized to low lying eigenstates of unbounded
hermitian operators\cite{KMDB80}.

The iteration scheme is then the following.  From $G_{\epsilon}V|n>$ calculate a normalized vector orthogonal to the vector $|n>$ 
\begin{equation}
|n_{\perp}>= (G_{\epsilon}V|n>-<n|G_{\epsilon}V|n>|n>)*\Vert\ldots\Vert^{-1}
\end{equation}
such that $<n_{\bot}|n_{\bot}>=1$, $<n|n_{\perp}>=0$ and $\Vert\ldots\Vert$ is the $\mathcal{L}^2$ norm. The matrix
representation of $G_{\epsilon}V$ in the subspace spanned by $|n>$
and $|n_{\bot}>$ is

\begin{equation}
(\begin{array}{cc}
\epsilon_{n} & v_{n}\\
v_{n}^{*} & \alpha_{n}
\end{array})\label{mx}
\end{equation} 
with
\begin{equation}
\epsilon_{n}=<n|G_{\epsilon}V|n>=\lambda_{n}^{-1},\; v_{n}=<n|G_{\epsilon}V|n_{\perp}>,\;\alpha_{n}=<n_{\perp}|G_{\epsilon}V|n_{\perp}>\label{lame}
\end{equation}

Now the 2x2 matrix (\ref{mx}) can be diagonalized yielding two eigenvalues
and the corresponding two eigenvectors. One of the eigenvalues always
lies above $\epsilon_{n}$ the other one below $\epsilon_{n}$ as
long as $v_{n}\neq0.$ (Here we make use of the fact, that $\lambda$ is real.) If we always chose the upper eigenvalue, the
$\epsilon_{n}$ form a monotonically increasing sequence bounded by
the highest eigenvalue of $G_{\epsilon}V$. Therefore the sequence
is convergent. (Similar arguments hold, if one chooses the lower eigenvalue
in each step in case the spectrum is bounded from below.) Since in
each step the modified iteration step is an optimization with respect
to the original power method step, one might expect an improved convergence
rate.
 
\begin{figure}
\centering
\begin{center}
\includegraphics[scale=0.7,angle=0.]{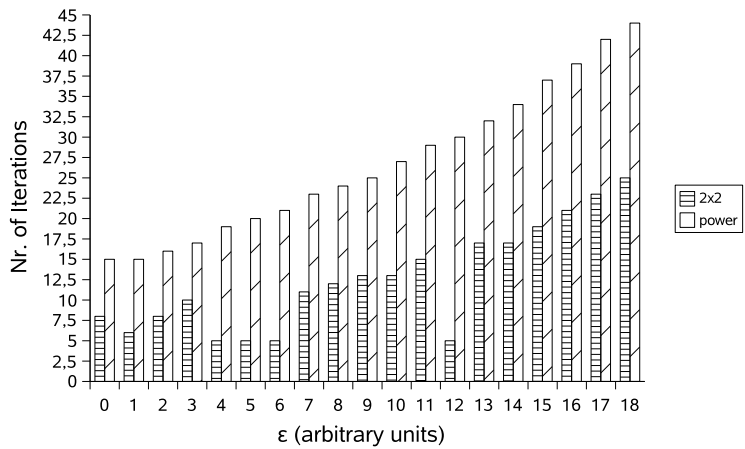}
\end{center}
\label{Fig1}\caption{Typical convergence rates of the coupling constant $\lambda$ for different ground state energies $\epsilon$ for Waxman algorithm (power) and the modified algorithm (2x2) using a 20x20 matrix model Hamiltonian.
 }
\end{figure}
\begin{figure}
\centering
\begin{center}
\includegraphics[scale=0.7,angle=0.]{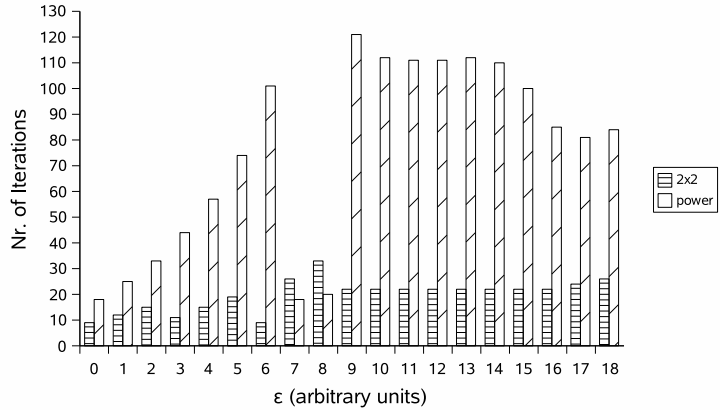}
\end{center}
\label{Fig2}\caption{ Convergence rates of the coupling constant $\lambda$ of different ground state energies $\epsilon$ for the Waxman algorithm (power) and the modified algorithm (2x2) using a 20x20 matrix which yields very poor convergence as model Hamiltonian.
 }
\end{figure}
In order to investigate the convergence properties we have performed
a number of calculations with matrices where a discrete spectrum
for $T$ was chosen and a random potential $V$ was added.
The signs in equation (\ref{heq}) were chosen such that $\lambda$ is positive
and the monotonically increasing iteration scheme (i.e. increasing in terms of  $\epsilon_{n}=\lambda_{n}^{-1}$ and therefore decreasing in terms of $\lambda_{n}$)     described above  was chosen. Figure 1 shows a typical convergence pattern for a 20x20 matrix using the original Waxman method (power) compared to the modified scheme proposed here (2x2). The number of iterations needed until a specified convergence limit is achieved for the new scheme is roughly one half that required for the original Waxman algorithm. Figure 2 illustrates the convergence properties of the modified iteration scheme  in another, more dramatic, case where the original algorithm has  considerable difficulty to converge. It can be seen from both examples, that the convergence rate improves significantly.

To judge the advantage of the faster convergence one has to take into account that because of the last term in equations (\ref{lame}) the new scheme is always one additional iteration step ($G_{\epsilon}V|n_{\perp}>)$ ahead and in each iteration step there are more vector operations. While the latter are negligible as far
as computational ressources are concerned, the additional iteration
step has to be taken into account. Even then this increases the
total number of iterations only by one and still results in a  considerably
accelerated convergence.

\begin{figure}
\centering
\begin{center}
\includegraphics[scale=0.7,angle=0.]{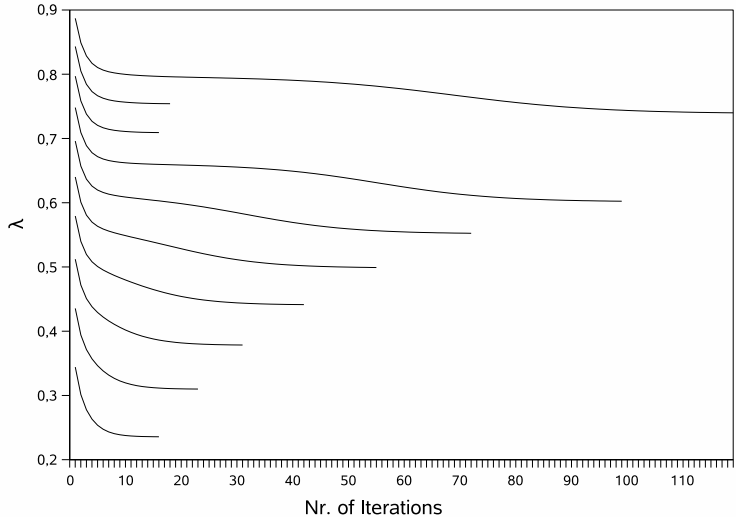}
\end{center}
\label{Fig3}\caption{Convergence curves of the coupling constant $\lambda$ for different ground state energies $\epsilon$ for Waxman algorithm using a 20x20 matrix as model Hamiltonian which causes psuedoconvergence.
 }
\end{figure}
\begin{figure}
\centering
\begin{center}
\includegraphics[scale=0.7,angle=0.]{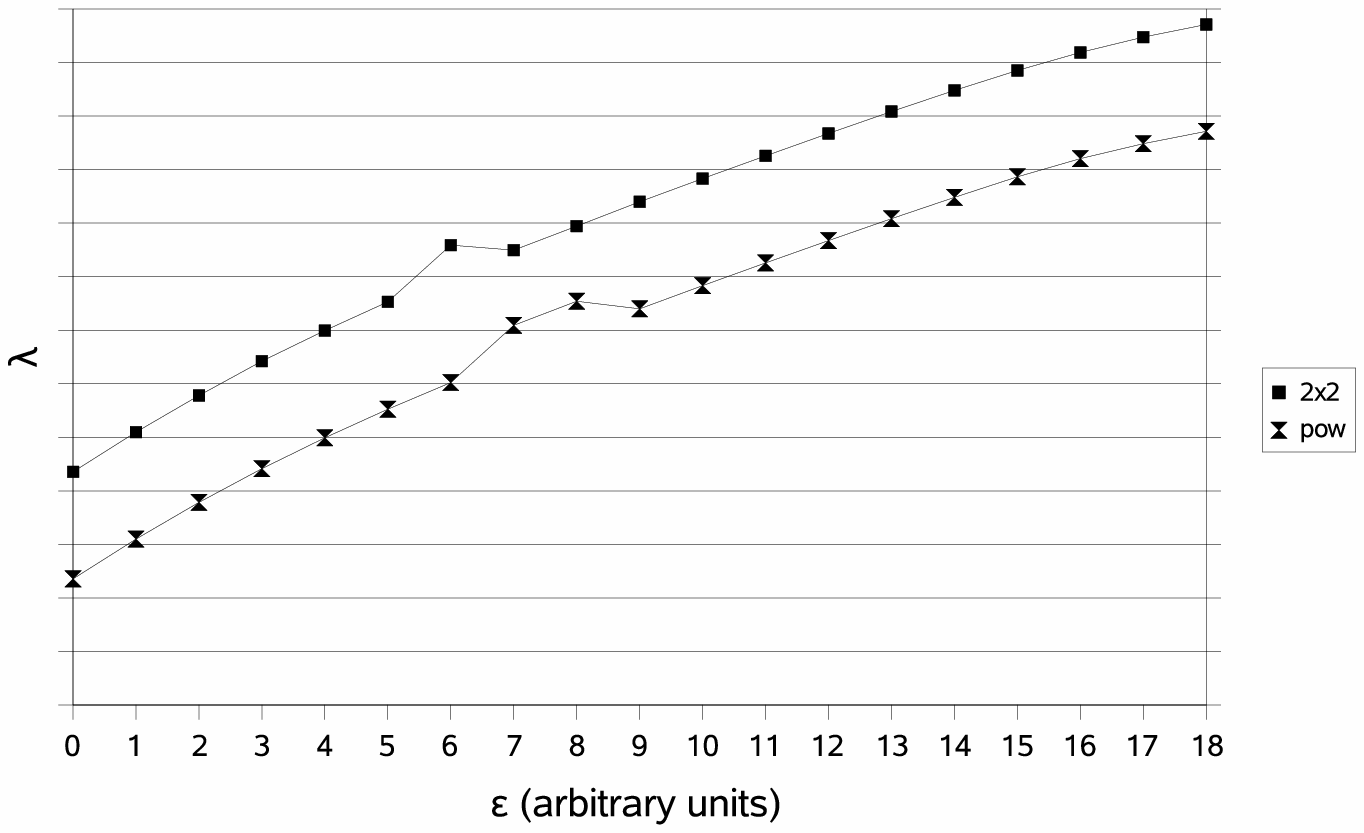}
\end{center}
\label{Fig4}\caption{Dependence of the numerically determined coupling constant $\lambda$ on the ground state energy $\epsilon$ for Waxman algorithm (power) and the modified algorithm (2x2) using a 20x20 matrix as model Hamiltonian which causes pseudoconvergence. The two curves which are identical when full convergence is achieved have been vertically shifted apart from each other for better visibility. 
 }
\end{figure}
 It is interesting to understand why achieving convergence in the second example is so slow. In Figure 3 we show the convergence of $\lambda_{n}$ for different energies as a function of iteration number n.  It can be seen, that convergence in certain cases is delayed by a number of steps in a such a way that seems to indicate that $\epsilon$ (or $\lambda$) appears   to converge to an incorrect value. Such behavior, which we is referred to as pseudoconvergence is common in several iterative algorithms and has been  discussed earlier for the Lanczos and modified Lanczos algorithm \cite{MB79}.

In general two problems arise with pseudoconvergence: How to detect and how to remedy it? The remedy in the present case may be to switch from the power to the 2x2 scheme. Since the 2x2 scheme itself suffers from pseudo convergence, further measures as discussed in \cite{MB79} may be appropriate.

The task of detecting pseudo convergence is more difficult since if you look e.g. at the rate of change of the eigenvalue in the successive iteration steps this can only indicate that there \textbf{may} be convergence. However, as has been shown recently for a wide class of potentials  smooth  relations between $\epsilon$ and $\lambda$ exist \cite{BM06}. We therefore show the corresponding relations for the pseudo convergence  exhibited in Figure 4. It can be seen, that the power method  exhibits marked  deviations from the smooth behavior at $\epsilon$-values 7 and 8. In Figure 2 we can see, that for these values the power method seemed to have  converged rapidly. In fact, and we checked this with the exact results,  pseudoconvergence occurred which   the  algorithm had failed to detect. Similarly in Figure 4 in the 2x2 scheme  one would  suspect that  pseudoconvergence occurred for $\epsilon=6$ which is actually  the case.

To summarize, the examples suggest that the modified algorithm has considerably improved convergence rates in general. In the case  where pseudoconvergence occurs the savings may become dramatic. The use of the Waxman algorithm enables one to detect the occurrence pseudoconvergence reasonably quickly from the marked deviations from the smooth dependence of $\lambda$ on $\epsilon$. As in  the power method scheme, the modified scheme does not require an explicit 
matrix representation in a large basis. Thus a major advantage of Waxman's method, namely the fact that the scheme can be applied directly to operators and wavefunctions in coordinate space, either using a discretized numerical or a parametrized analytical representation \cite{BM06}, is preserved when migrating to the modified algorithm.


\end{document}